\documentclass[12pt]{article}
\usepackage{epsfig}
\usepackage[hang,bf,small]{caption}
\DeclareGraphicsExtensions{.eps.gz,.eps,.ps,.ps.gz}
\parskip=\itemsep               %?

\newcommand{\beq}{\begin{equation}}
\newcommand{\eeq}{\end{equation}}
\newcommand{\beqar}[1]{\begin{eqnarray}\label{#1}}
\newcommand{\eeqar}{\end{eqnarray}}

\newcommand{\si}{\sigma}

\newcommand{\as}{\alpha_S}

\def\eq#1{{Eq.~(\ref{#1})}}

\def\plb#1#2#3{    {\it Phys. Lett. }{\bf B#1} (19#2) #3}
\def\prd#1#2#3{    {\it Phys. Rev. }{\bf D#1} (19#2) #3}

\def\zpc#1#2#3{    {\it Z. Phys. }{\bf C#1} (19#2) #3}

\relax
\begin{document}
\title{
{\Large \bf   Diffractive Dissociation  and Saturation Scale }\\\
{ \Large \bf  from Non-Linear Evolution  in  High Energy DIS}}
\author{
{\large  ~ E.~Levin\thanks{e-mail: 
leving@post.tau.ac.il}~~$\mathbf{{}^{a),b)}}$
 \,~\,and\,\,~
M. ~Lublinsky\thanks{e-mail: 
mal@techunix.technion.ac.il}~~$\mathbf{{}^{c),d)}}$}\\[2.5ex]
{\it ${}^{a)}$ \small HEP Department}\\
{\it \small  School of Physics and Astronomy}\\
{\it\small  Raymond and Beverly Sackler Faculty of Exact Science}\\
{\it \small Tel Aviv University, Tel Aviv 69978, ISRAEL}\\[2.5ex]
{\it ${}^{b)}$\small  DESY Theory Group,}\\
{ \it\small  D-22602, Hamburg, GERMANY}\\[2.5ex]
{\it ${}^{c)}$ \small  Department of Physics}\\
{\it \small  Technion -- Israel Institute of   Technology}\\
{\it \small  Haifa 32000, ISRAEL}\\[2.5ex]
{\it ${}^{d)}$ \small  II. Institut f\"{u}r Theoretische Physik,  Universit\"{a}t Hamburg}\\
{\it\small  Luruper Chaussee 149, 
       22761 Hamburg, GERMANY  }\\[1.5ex]
}

\maketitle
\thispagestyle{empty}
                      
\begin{abstract} 
This paper presents the first numerical solution to the non-linear 
evolution equation for diffractive 
dissociation processes in deep inelastic scattering.
It is shown that the solution depends on one scaling variable 
$\tau = Q^2/Q^{D\,2}_s(x,x_0)$, 
where $Q^D_s(x,x_0)$ is the saturation scale for the diffraction 
processes. The  dependence of the saturation scale 
$Q^D_s(x,x_0)$ on both $x$ and $x_0$ is  investigated, ($Y_0 = \ln(1/x_0)$ is a 
minimal rapidity gap for the diffraction process). 
The $x$ - dependence of $Q^D_s$ turns out to be the same as of the saturation scale in 
the total inclusive DIS cross section. In our calculations $Q^D_s(x,x_0)$ reveals only 
mild dependence on $x_0$. The scaling is shown to hold for $x \ll x_0$ but  is violated 
at $ x \sim x_0$. 

 \end{abstract}
\thispagestyle{empty}
\begin{flushright}
\vspace{-21.5cm}
DESY-01-122 \\
TAUP - 2687 - 2001 \\
\today
\end{flushright}   
\newpage
\setcounter{page}{1}

\section{Introduction}
\setcounter{equation}{0}

Diffractive inclusive production in deep inelastic scattering (DIS) at 
high energy  has become 
an area of  particular interest of experts since it provides a deeper 
insight into dynamics of QCD in the kinematic region where the density of 
partons  is expected to be high (see Ref.\cite{WM} and reference therein).

Inclusive diffraction in DIS offers an opportunity to probe the 
transition region between ``soft" and ``hard" interactions giving  
natural estimates for the value of the shadowing corrections in DIS, 
namely $ \Delta F_2 = F_2 - F_2^{DGLAP} =  F_D $ \footnote{$F_D$ is 
the diffractive structure 
function introduced in Ref. \cite{ZEUSDATA}.}
 as was firstly shown in Ref. \cite{LW} on the 
basis of the AGK cutting rules \cite{AGK}.  
A more detail approach  started with the 
Kovchegov-McLerran\cite{KM}  
formula which expresses the ratio of the diffraction cross section 
($\sigma_{diff}$) to the total cross section ($\sigma_{tot}$) 
in DIS initiated by the 
quark-antiquark pair produced in 
$\gamma^* \,\rightarrow\,q\,\,+\,\,\bar q $  decay of the virtual 
photon.  This formula reads 

\beq \label{KMF}
R\,\,=\,\,\frac{\sigma_{diff}}{\sigma_{tot}}\,\,=\,\,\frac{\int\,d^2\,b \int
\,dz \int \,d^2  r_{\perp}
P^{\gamma^*}(z,r_{\perp};Q^2)\, N^2(r_{\perp},x;b)}
{ 2\,  \int\,d^2 b \,\int\,dz
\,\int\,d^2\,r_{\perp}\,\,P^{\gamma^*}(z,r_{\perp};Q^2)\,\,
N ( r_{\perp},x;b)
 }\,\,.
\eeq
where $N(r_{\perp},x;b)$ is the imaginary part of 
the elastic dipole-target 
amplitude for dipole of the size $r_\perp$ scattered at fixed Bjorken 
$ x = 
Q^2/W^2$ ($ Q^2$ is the photon virtuality and $W$ is its energy in the target rest frame) and at 
fixed impact $b$.  
$P^{\gamma^*}(z,r_{\perp};Q^2)$ is the probability to find a 
quark-antiquark pair with size $r_{\perp}$ inside the virtual photon 
\cite{MU90,WF}:
\begin{eqnarray}
P^{\gamma^*}(z,r_{\perp};Q^2)&=&\frac{\alpha_{em} N_c}{2
\pi^2}
\,\sum_f \,Z^2_f \sum_{\lambda_1,\lambda_2}\,\{\, | \Psi_T |^2\,\,+\,\,|
\Psi_L|^2 \,\}\,\,\label{PROBPH}\\
&=&\frac{\alpha_{em} N_c}{2 \pi^2}
\sum_f Z^2_f \,\{\,( z^2 + ( 1 - z )^2 )a^2 K^2_1( a\,r_{\perp}
)\,+\,4\,Q^2\,z^2( 1 - z )^2 K^2_0(
a\,r_{\perp})\,\},\nonumber
\end{eqnarray}
where $a^2 = z(1-z)Q^2 + m^2_q$. The functions $\Psi_{T,L}$ stand for transverse  and
longitudinal polarized photon wave functions. 
\eq{KMF} is important since it provides a relation 
between the dipole-target elastic amplitude and the 
cross section of the diffraction dissociation. A non-linear 
evolution equation was derived 
for the former \cite{GLR,MUQI,MU94,BA,KO,Braun,ILM}. 
This equation  has been studied both 
analytically \cite{ILM,LT} and numerically \cite{Braun,LGLM,LL,Braun2}. 

The formula (\ref{KMF}) fails to describe correctly the experimental data on 
the diffraction production. Moreover, inclusion of an extra gluon emission 
in the initial virtual photon wave function is still 
insufficient to reproduce 
the data
 \cite{GLMDD,KOP,KOVN,KOV}.
Nevertheless, \eq{KMF} can be viewed rather as  initial 
condition to a more complicated  equation.

The non-linear equation for the diffraction dissociation 
processes can be written for the amplitude $N^D$ which has the following
meaning \cite{LK}.

We introduce  the cross section for diffraction production  with the 
 rapidity gap larger than given $Y_0\equiv\ln (1/x_0)$:
\beq
\label{F2D}
\si_{diff}(x,x_0,Q^2)\,\,\,=\,\,\int\,\,d^2 r_{\perp} \int \,d
z\,\,P^{\gamma^*}(z,r_{\perp};Q^2)
 \,\,\sigma_{\rm dipole}^{diff}(r_{\perp},
x,x_0)\,,
\eeq 
  and 
\beq \label{DDCX}
\sigma_{\rm dipole}^{diff}(r_{\perp},x,x_0) \,\,=\,\,\int\,d^2
b\,\,N^D(r_{\perp},x,x_0;b)\,.
\eeq

The function $N^D$ is the amplitude
of the diffraction production induced by the dipole 
with size $r_{\perp}$  with rapidity
gap larger than given ($Y_0$). 
Note that the minimal rapidity gap $Y_0$ can be kinematically
 related to the maximal  diffractively  produced mass: $x_0=(Q^2+M^2)/W^2$. 

The non-linear evolution equation for  $N^D$ was derived in Ref. \cite{LK}
and recently rederived in Ref. \cite{KOVN}:
\begin{eqnarray} 
   N^D({\mathbf{x_{01}}},Y,Y_0;b)  =  N^2({\mathbf{x_{01}}},Y_0;b)\, 
{\rm e}^{-\frac{4
C_F\,\as}{\pi} \,\ln\left( \frac{{\mathbf{x_{01}}}}{\rho}\right)(Y-Y_0)}\,
+\frac{C_F\,\as}{\pi^2}\int_{Y_0}^Y dy \,  {\rm e}^{-\frac{4
C_F\,\as}{\pi} \,\ln\left( \frac{{\mathbf{x_{01}}}}{\rho}\right)(Y-y)} \times  \nonumber \\
\nonumber \\
 \int_{\rho} \, d^2 {\mathbf{x_{2}}}  
\frac{{\mathbf{x^2_{01}}}}{{\mathbf{x^2_{02}}}\,
{\mathbf{x^2_{12}}}} 
[\,2\,  N^D({\mathbf{x_{02}}},y,Y_0;{ \mathbf{ b-
\frac{1}{2}
x_{12}}})
+  N^D({\mathbf{x_{02}}},y,Y_0;{ \mathbf{ b - \frac{1}{2}
x_{12}}})  N^D({\mathbf{x_{12}}},y,Y_0;{ \mathbf{ b- \frac{1}{2}
x_{02}}}) \nonumber \\ \nonumber \\
- 4 \, N^D({\mathbf{x_{02}}},y,Y_0;{ \mathbf{ b - \frac{1}{2}
x_{12}}})  N({\mathbf{x_{12}}},y;{ \mathbf{ b- \frac{1}{2}
x_{02}}})+2\, N({\mathbf{x_{02}}},y;{ \mathbf{ b -
\frac{1}{2}
x_{12}}})  N({\mathbf{x_{12}}},y;{ \mathbf{ b- \frac{1}{2}
x_{02}}})
]\,. \nonumber \\ \label{DDEQ}
\end{eqnarray}
The evolution (\ref{DDEQ}) is a subject to initial conditions at $x=x_0$:
\beq\label{iniDD}
N^D(r_\perp,x_0,x_0;b)\,=\,N^2(r_\perp,x_0;b)\,.
\eeq

Namely, at the energy equal to the energy gap diffraction is purely given 
by the elastic scattering as 
it was stated in \eq{KMF}.  

Since at high energies color dipoles are correct  degrees of freedom
\cite{MU94}  we can write  the 
unitarity  constraint :
\beq\label{F}
2\,N\,=\,N^D\,+\,F\,,
\eeq
where the function $F$ denotes contributions of all the inelastic processes.  An important
observation is that $F$ satisfies the same equation as $N$ \cite{BA,KO} but with shifted initial 
conditions \cite{LK}:
\beq\label{iniF}
F_{ini}\,=\, N_{ini}\,-\,N^2_{ini}
\eeq

Another interesting quantity to study is the cross section of diffractive dissociation process
with a fixed gap or equivalently to a fixed mass:
\beq\label{R}
\Re\,\equiv \,-\,\partial N^D/\partial Y_0\,.
\eeq
The function $\Re$ was introduced in Ref. \cite{LK}. The authors of this paper proposed
a model in which $\Re$ was shown to possess a maximum when varying  $Y_0$
at fixed $Y$. Physically this maximum means that at given $Y$ there is a preferable mass
for the production. Below we will argue that the appearance of the maximum is related 
to the scaling phenomena to be displayed by the function $N^D$.

The present paper is entirely  devoted to the numerical solution of the equation (\ref{DDEQ}). Various
properties of the solutions $N^D$ are investigated while our final goal computation of the diffraction 
cross section will be published  separately \cite{LL1}. 
In the next Section (2) the solution of the equation  
(\ref{DDEQ}) is presented. Section 3 deals with the determination of the diffractive saturation scale.
Scaling phenomena is discussed in  Section 4. We conclude in the last Section (5).

\section{Solution of the non-linear equation}

In this section we report on the numerical solution of the equation (\ref{DDEQ}). The method
of iterations proposed in Ref. \cite{LGLM} is applied. The constant value for the strong 
coupling constant $\as=0.25$ is always used.The solutions are computed for 
$4\times 10^{-5} \le x_0\le 10^{-2}$ and within the kinematic region $10^{-7}\le x \le x_0$ 
and distances up to a few fermi.

The function $N^D$ is formally a function of four variables: the energy gap $x_0$,
 the Bjorken variable $x$, the transverse
distance $r_\perp$, and the impact parameter $b$. The $b$-dependence is  parametric only 
because the  evolution kernel does not depend on $b$. In order to simplify the problem
we will proceed similarly to the treatment of the $b$-dependence of the function $N$
\cite{LGLM}. In that paper  we assumed the function $N$   to 
preserve the very same $b$-dependence as introduced 
in the initial conditions:
\beq
\label{Nb}
 N(r_\perp,x; b)\,=\, (1\,-\,e^{-\kappa(x,r_\perp)\, S(b)}),
\eeq
with the function $\kappa$ being related to the ``$b=0$'' solution $\tilde N(r_\perp,x)$:
\beq
\label{kappa}
\kappa(x,r_\perp)\,=\,-\,\ln(1\,-\,\tilde N(r_\perp,x)).
\eeq
$\tilde N(r_\perp,x)$ represents a solution of the very same non-linear equation (see 
Refs.\cite{BA,KO})  but with
no dependence on the third variable. The initial conditions for the function 
$\tilde N(r_\perp,x)$ are taken at $b=0$. For the case of the proton target \cite{LGLM} 
the anzatz in the form (\ref{Nb}) was shown to be a quite  good approximation of the exact
$b$-dependence of the solution to the non-linear equation for $ N(r_\perp,x;b)$. In Ref. 
\cite{LL} we investigated the anzatz
(\ref{Nb}) for the gold target and again found it to be a very good approximation at least
for impact parameters smaller than the target radius.

In order to be consistent with  initial conditions (\ref{iniDD}) we assume the following
$b$-dependence of $N^D$:
\beq
\label{NDb}
 N^D(r_\perp,x,x_0; b)\,=\, (1\,-\,e^{-\kappa^D(x,x_0,r_\perp)\, S(b)})^2,
\eeq
with 
\beq
\label{kappaD}
\kappa^D(x,x_0,r_\perp)\,=\,-\,\ln(1\,-\,\sqrt{\tilde N^D(r_\perp,x,x_0)}).
\eeq
$\tilde N^D(r_\perp,x,x_0)$ represents a solution of the  equation (\ref{DDEQ}) 
but with no dependence on the forth variable. The initial conditions for the function 
$\tilde N^D(r_\perp,x,x_0)$ are set at $b=0$ and  $\kappa^D(x_0,x_0,r_\perp)\,=\,
\kappa(x_0,r_\perp)$. Since in the present paper we do not intend to compute cross sections,
for which we would need to perform the $b$ integration, the accuracy of the anzatz (\ref{NDb})
will not be investigated here.

\begin{figure}[htbp]
\begin{tabular}{c c c}
$x=10^{-7}$ & $x=10^{-5}$ & $x=10^{-3}$ \\ 
 \epsfig{file=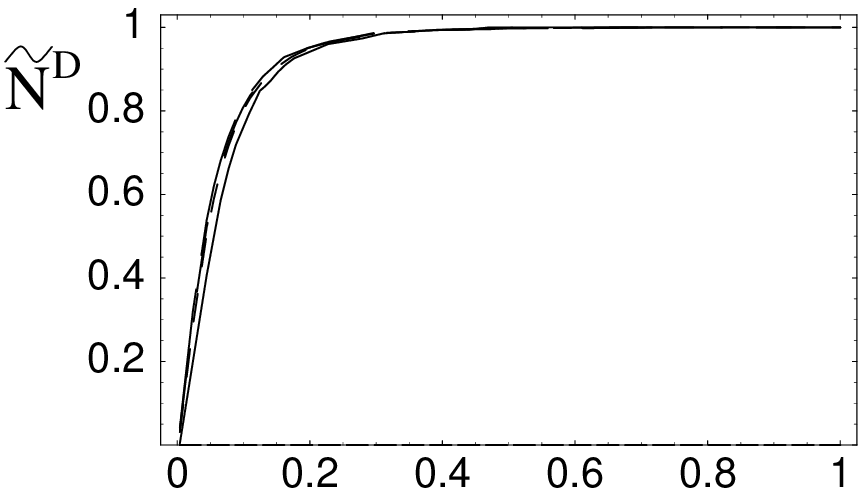,width=54mm, height=42mm}&
\epsfig{file=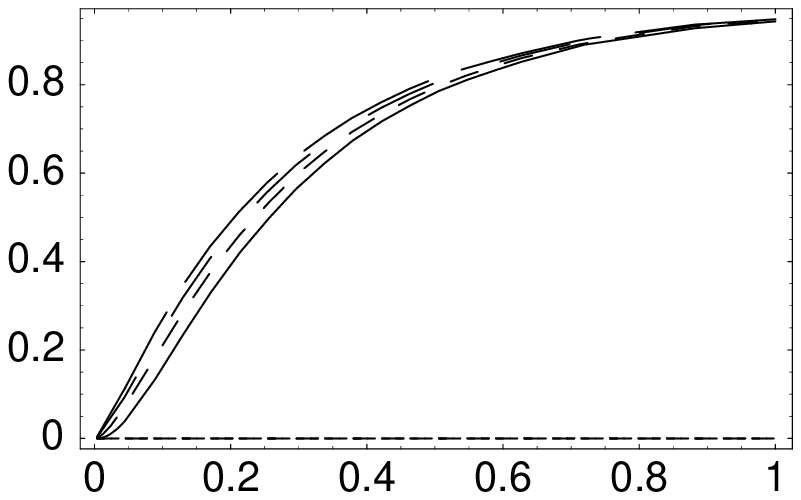,width=48mm, height=42mm}&  
 \epsfig{file=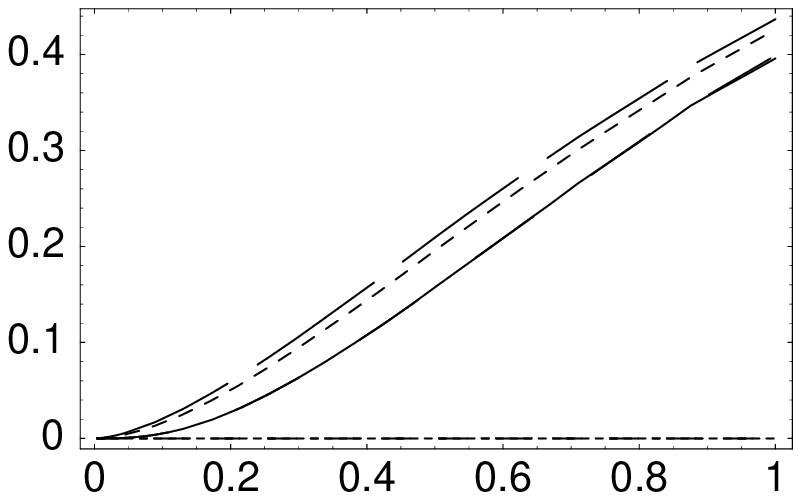,width=48mm, height=42mm} \\
\epsfig{file=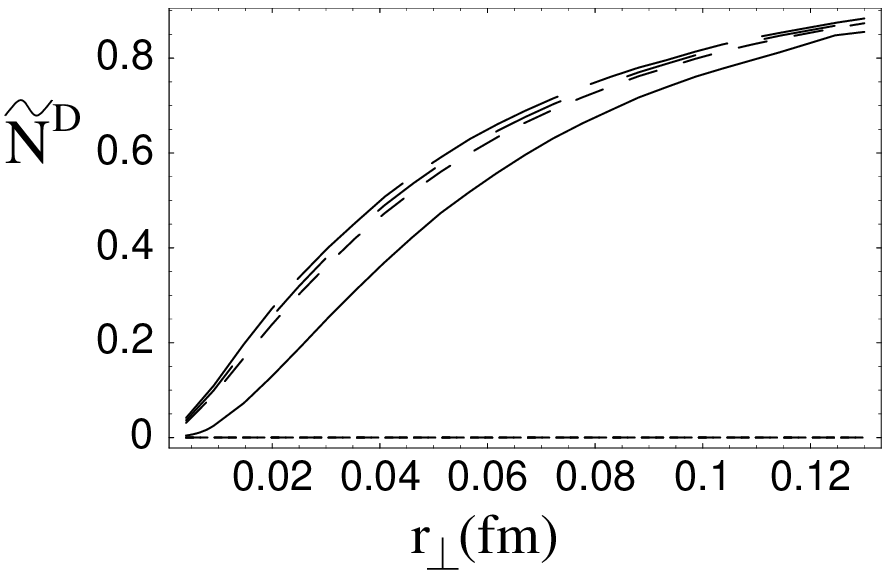,width=54mm, height=45mm}&
 \epsfig{file=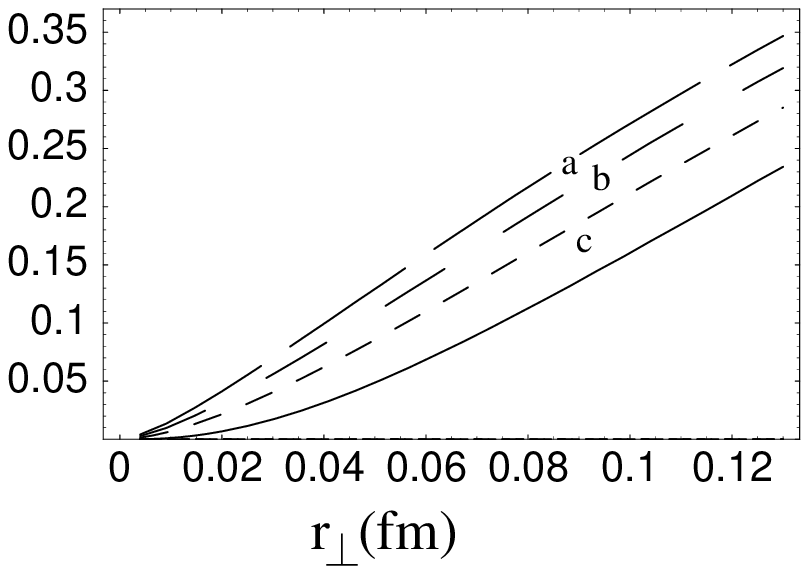,width=48mm, height=45mm}&
\epsfig{file=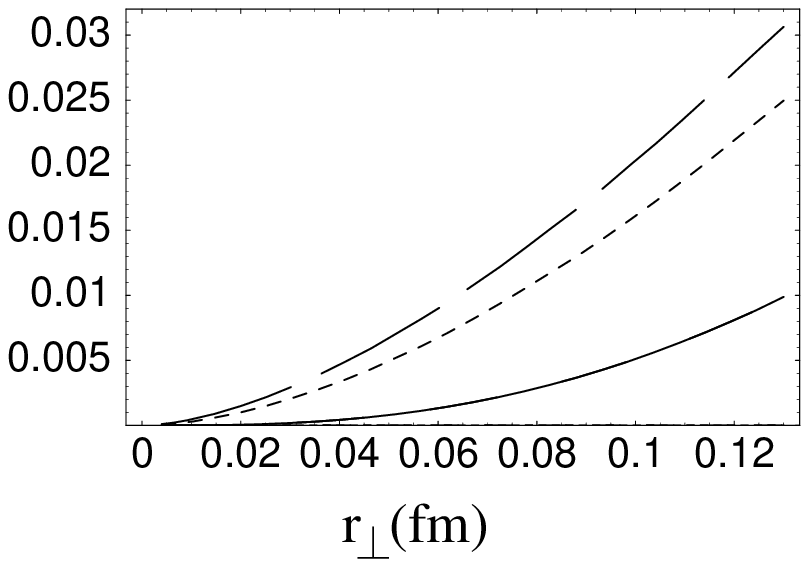,width=48mm, height=45mm}\\
\end{tabular}
  \caption[]{\it The function $\tilde N^D$ is plotted versus distance. The curves correspond
to different values of $x_0$: a - $x_0=10^{-2}$;  b - $x_0=10^{-3}$; c - $x_0=10^{-4}$.
 The solid line is $\tilde N^2$.}
\label{solution}
\end{figure}

For each initial value of $x_0$  the function $\tilde N^D(r_\perp,x,x_0)$ 
is obtained after about ten iterations. 
The Fig.~\ref{solution} shows the solutions $\tilde N^D$ as a function of the distance 
for various values of  $x_0$ and $x$. The amplitude
for the elastic scattering $\tilde N^2$ \cite{LGLM} is plotted in the same graph. 
The obtained numerical 
inequality $\tilde N^2\le \tilde N^D\le\tilde N$ is in perfect agreement with the physical expectations for
the diffractive dissociation cross section to be larger than the elastic cross section. Another
consistency check is the saturation of the function $\tilde N^D$ which is a consequence of
the unitarity bound.  In the black disk limit diffractive dissociation is a half of the total cross
section.

\begin{figure}[htbp]
\begin{tabular}{c c c c }
 \epsfig{file=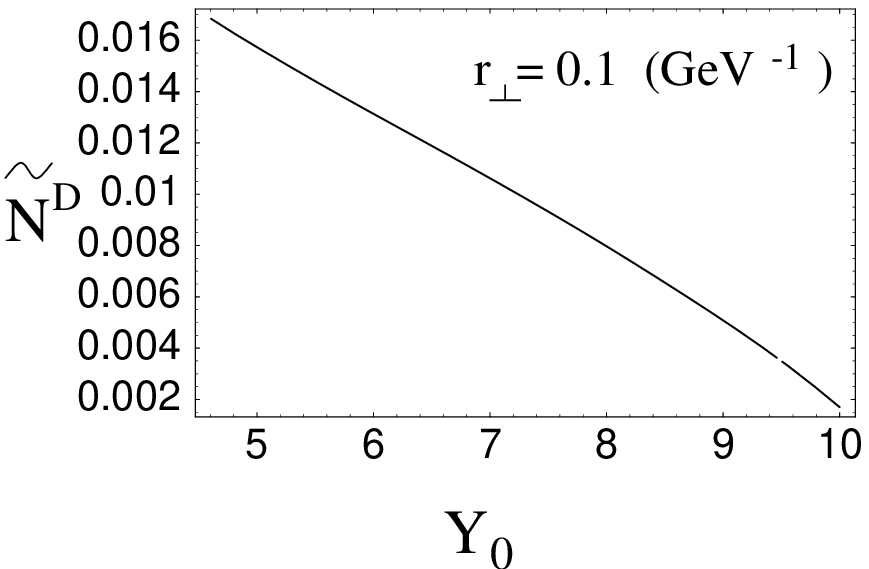,width=44mm, height=38mm}&
\epsfig{file=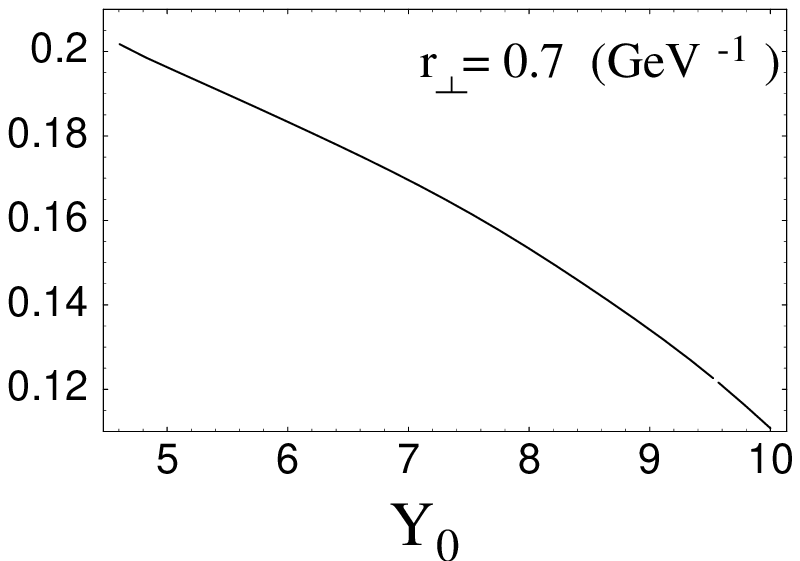,width=37mm, height=38mm}& 
 \epsfig{file=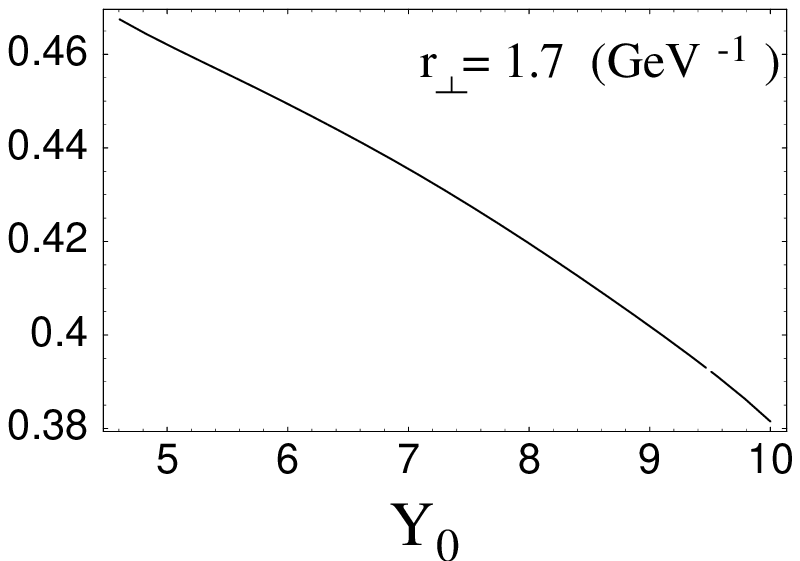,width=37mm, height=38mm} &
\epsfig{file=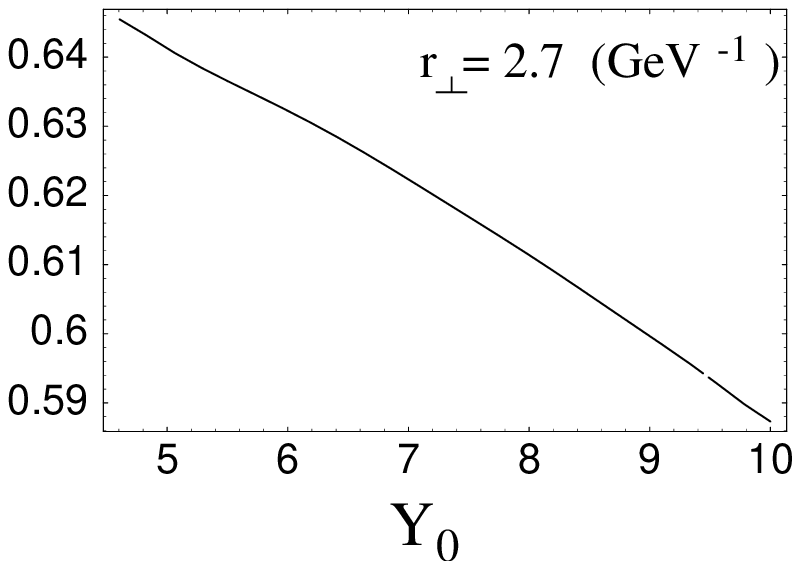,width=37mm, height=38mm}\\
\end{tabular}
  \caption[]{\it The function $\tilde N^D$ is plotted versus the gap $Y_0$ at fixed rapidity 
$Y=10$.}
\label{solution2}
\end{figure}

It is worth to investigate the dependence of the solutions obtained  on the gap
variable $x_0$. To this goal we plot the function $N^D$ as a function of the gap $Y_0$
for various distances and at fixed $Y=10$ (Fig.~\ref{solution2}). At short distances the solution 
depends strongly
 on $x_0$ though as we approach  the saturation region this dependence dies out.

It was stated in the Introduction that the function $N^D$ equals $2 N - F$, where both 
functions $N$ and $F$ are solutions of the same non-linear equation \cite{BA,KO}. Thus it is natural 
to compute
 $2 N - F$ solving  the non-linear evolution equation 
\cite{BA,KO}  with appropriate initial 
conditions.
 A comparison with $N^D$ from (\ref{DDEQ}) would be an
ultimate test for the correctness of the numerical  procedures.  Such test was successfully
performed and we found an absolute agreement (relative error less than 1\%)
between both the computations. 

\section{Saturation Scale}

Determination of the diffractive saturation scale $Q_s^D(x,x_0)$ from the solution $\tilde N^D$ 
is a subject of this section. Unfortunately, no exact mathematical 
definition of the saturation scale is known so far. In the Refs. \cite{LGLM,me} several definitions
of the saturation scale $Q_s(x)$ were proposed which related  saturation scale to the shape
of the  function $\tilde N$. It is important to stress that it is not clear a priori whether  $Q_s^D$
should coincide with $Q_s$ or not.
We will proceed here in the same spirit as Ref. \cite{me}. Namely,
we propose several definitions of the saturation scale while the variety of the obtained 
results will indicate the uncertainty in the definitions.  

For a step like function it is natural to define the saturation scale as position where 
it reaches half of the maximum:
\begin{itemize}
\item {\bf Definition (a):}
\beq \label{defa}
\tilde N^D(R_s^D,x,x_0)\,=\,1/2\,,\,\,\,\,\,\,\,\,\,\,\, Q_s^D\,\equiv\, 2/R_s^D\,.
\eeq
\end{itemize}
The equality between the saturation radius $R_s^D$ and the saturation scale 
$Q_s^D$  is motivated by the double logarithmic approximation. Though
 this approximation is formally  not justified, we still believe it to make reliable
estimates provided $Q_s^D$ is  large enough.
The definition (\ref{defa}) is analogous to the one proposed in Ref. \cite{LGLM} 
$N(2/Q_s,x)=1/2$. If we recall that  $N^D=N^2$ at  $x=x_0$ and postulate $Q_s^D(x_0,x_0)
=Q_s(x_0)$ then consistency requires
\begin{itemize}
\item {\bf Definition (b):}
\beq \label{defb}
\tilde N^D(2/Q_s^D,x,x_0)\,=\,1/4\,.
\eeq
\end{itemize}

An alternative definition of the saturation scale could be one motivated by the 
Glauber-Mueller formula:
\begin{itemize}
\item {\bf Definition (c):}
\beq \label{defc}
\kappa^D(2/Q_s^D,x,x_0)\,=\,1/2\,.
\eeq
\end{itemize}

\begin{figure}[htbp]
\begin{tabular}{c c c  }
 \epsfig{file=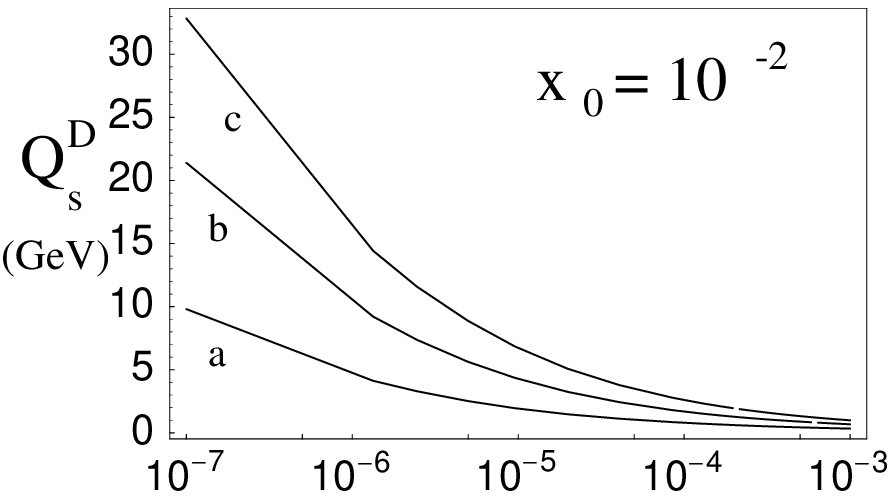,width=55mm, height=38mm}&
\epsfig{file=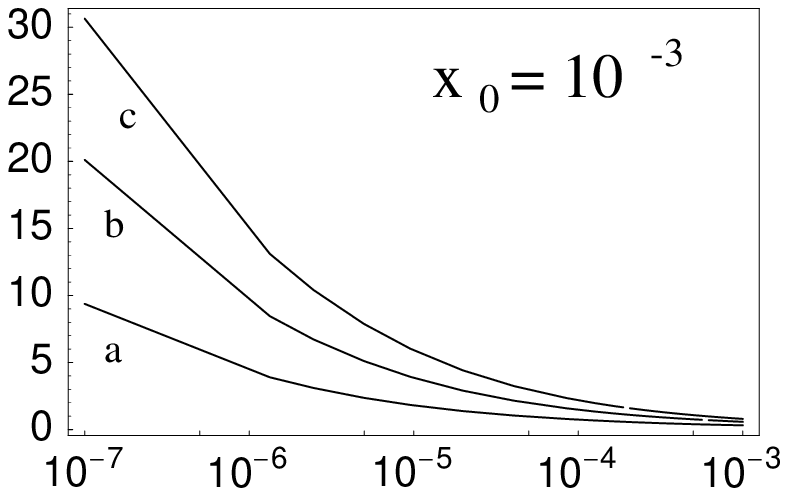,width=52mm, height=38mm}& 
 \epsfig{file=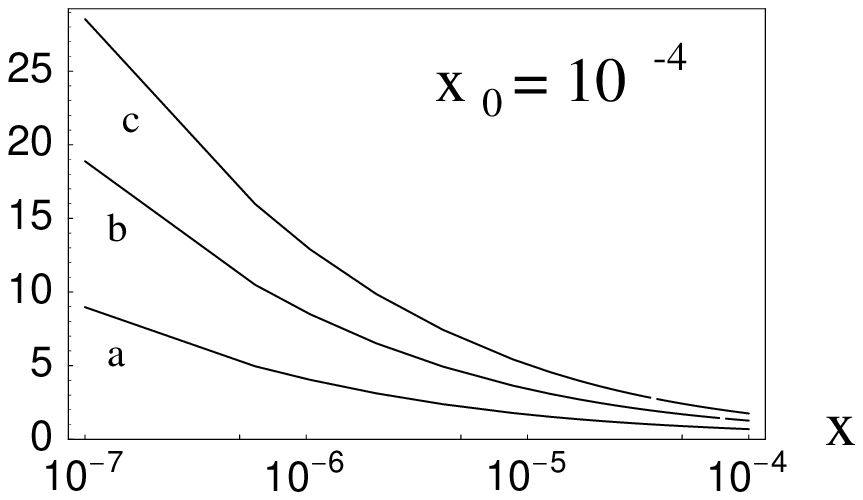,width=55mm, height=38mm} \\
\end{tabular}
  \caption[]{\it The saturation scale $Q_s^D$ is plotted versus $x$. The three curves correspond
to the definitions (\ref{defa}) (lowest curve),  (\ref{defb}) (middle curve), and  
(\ref{defc}) (upper curve).}
\label{Qs}
\end{figure}

The saturation scales deduced through the above definitions are depicted in Fig. \ref{Qs}.
For given $x_0$ the observed hierarchy 
between the saturation scales obtained is an obvious consequence of the definitions 
 (\ref{defa}),   (\ref{defb}),  (\ref{defc}) and the shape of the function $\tilde N^D$ 
(Fig. \ref{solution}). Note that the saturation scale is almost $x_0$ independent.

It is important to learn about $x$-dependence of the saturation scale. To this goal,
we assume the following parameterization:
\beq\label{qsat}
Q_s^D(x,x_0)\,=\,Q_{s\,0}^D\,\,x^{-\lambda}\,\,x_0^\beta\,.
\eeq
In fact, the parameterization (\ref{qsat}) is a good approximation for the values 
of the saturation scales obtained with
$$
\lambda\,=\,0.385\,\pm\,0.015\,; \,\,\,\,\,\,\, {\rm and}\,\,\,\,\,\, \beta\,=\,0.045\,\pm\,0.025\,.
$$
Within the errors these powers coincide for all the saturation scale definitions
 (\ref{defa}),   (\ref{defb}),  (\ref{defc}). The small value for the power $\beta$ is
a numeric  indication of the very weak $x_0$-dependence of the saturation scale. Its 
large relative error results on one hand from numerical limitations and  on the other hand,
this error signals for more complicated $x_0$-dependence than it is given in  (\ref{qsat}).

It is important to stress that the obtained power $\lambda$ coincides with the corresponding
power of the saturation scale $Q_s$ \cite{me}. 

\section{Scaling phenomena}

In the Ref. \cite{me} the function $\tilde N$ was shown to display the scaling phenomena.
We present here a similar analysis for the function $\tilde N^D$. 
In the saturation region the scaling  
implies the amplitude to be a function of only one variable $\tau= (r_\perp\cdot Q_s^D(x,x_0))^2$:
\beq\label{SCALING}
\tilde N^D(r_\perp,x,x_0)\,=\,\tilde N^D(\tau)\,.
\eeq

Let us define the following derivative functions assuming the 
scaling behavior (\ref{SCALING}):
\beq 
 N_y^D(r_\perp,x,x_0)\,\equiv\,-\,\frac{\partial \tilde N^D}{\partial Y}\,=\,
  \frac{d \tilde N^D}{d \tau}\,\tau\,\frac{\partial\ln (Q_s^D)^2}{\partial \ln x}\,,
\label{DX}
\eeq
\beq
 N_r^D(r_\perp,x,x_0)\,\equiv\, r_\perp^2\,\frac{\partial \tilde N^D}{\partial r_\perp^2}\, =\,
 \frac{d \tilde N^D}{d \tau}\,\tau\,,
\label{DR}
\eeq
\beq
 \Re(r_\perp,x,x_0)\,\equiv\,-\frac{\partial \tilde N^D}{\partial Y_0}\,=\,
  \frac{d \tilde N^D}{d \tau}\,\tau\,\frac{\partial\ln (Q_s^D)^2}{\partial \ln x_0}\,.
\label{DX0}
\eeq

\begin{figure}[htbp]
\begin{tabular}{c c c c}
 \epsfig{file=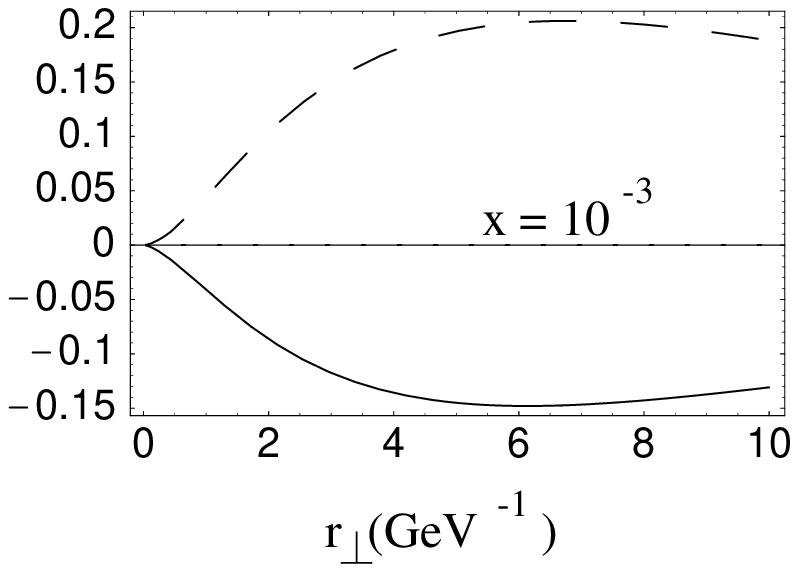,width=40mm, height=40mm}&
\epsfig{file=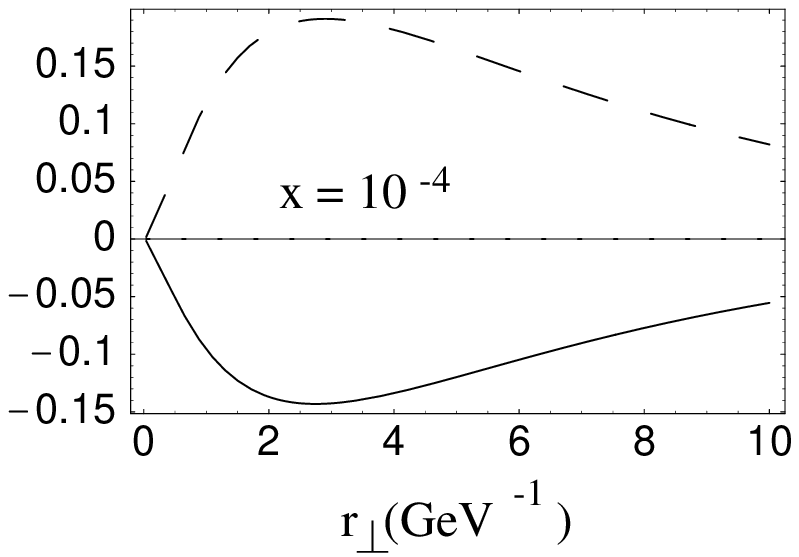,width=40mm, height=40mm}&
 \epsfig{file=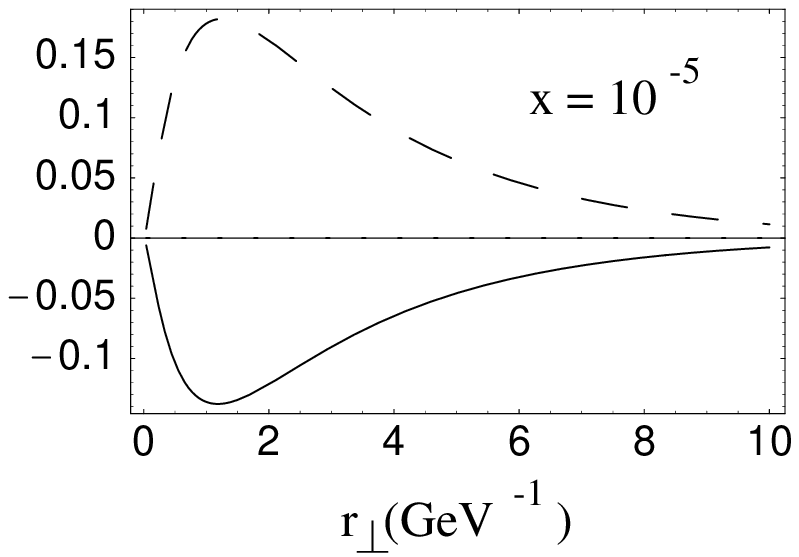,width=40mm, height=40mm}&
\epsfig{file=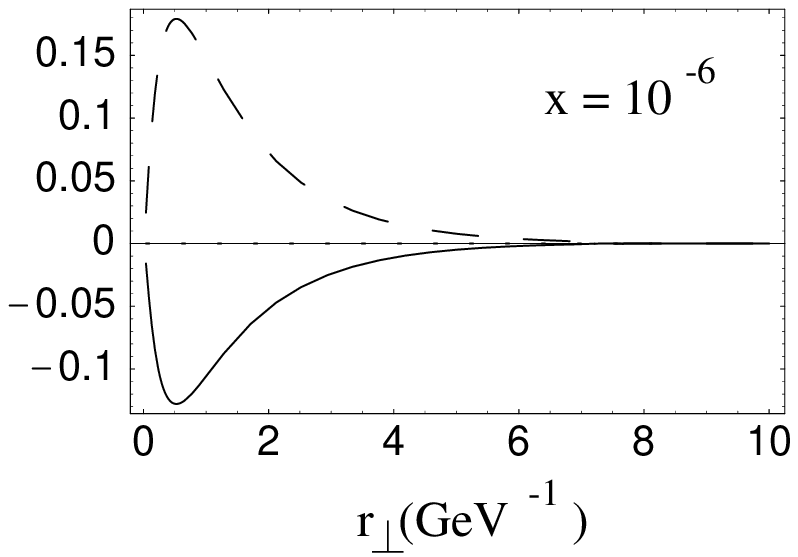,width=40mm, height=40mm}\\ 
\end{tabular}
  \caption[]{\it The derivative functions $N_r^D$ (dashed line) and $N_y^D$ (solid line) 
as functions of the distance  at fixed $x_0=10^{-2}$. }
\label{scal_x}
\end{figure}

\begin{figure}[htbp]
\begin{tabular}{c c c c}
 \epsfig{file=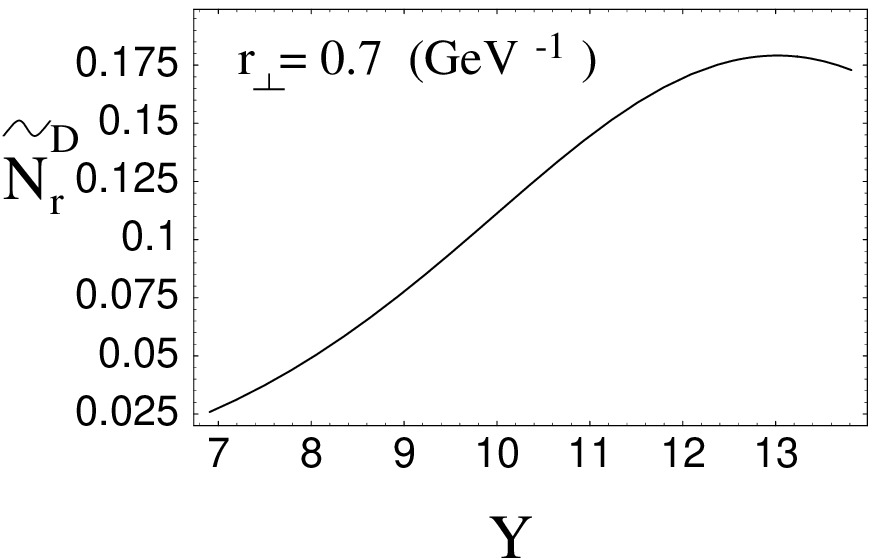,width=42mm, height=40mm}&
\epsfig{file=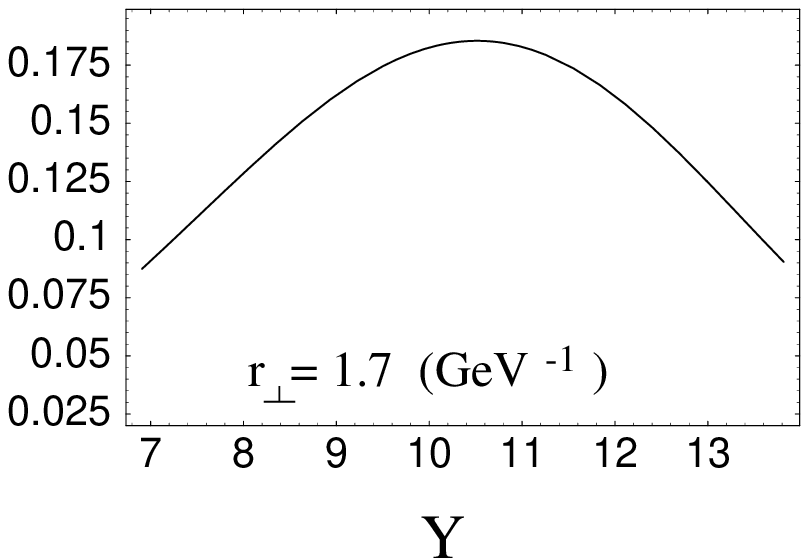,width=38mm, height=40mm}&
 \epsfig{file=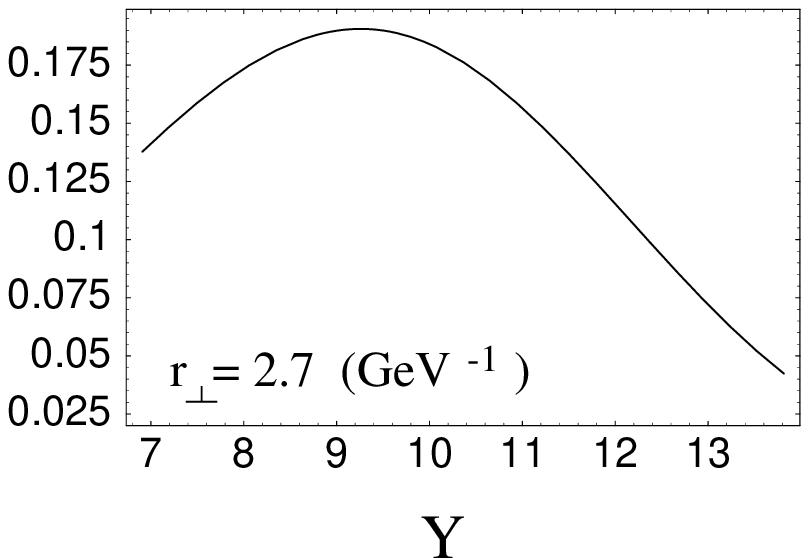,width=38mm, height=40mm}&
\epsfig{file=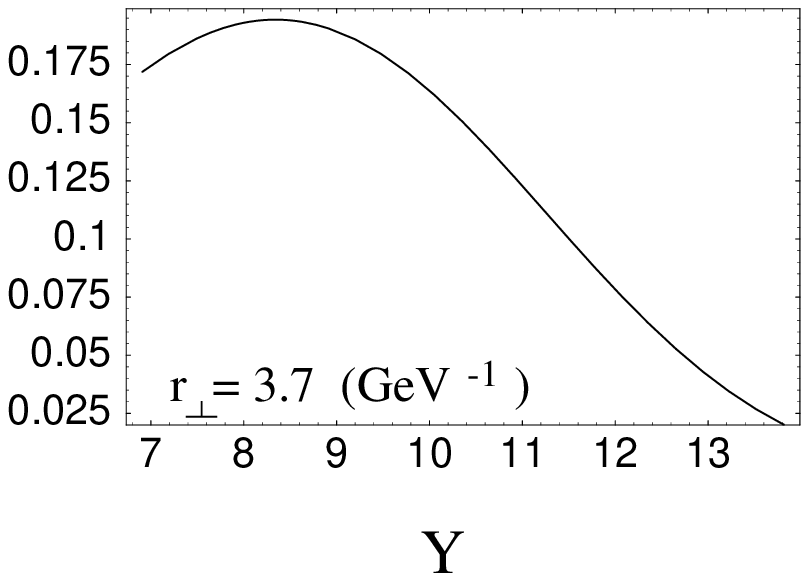,width=38mm, height=40mm}\\ 
\end{tabular}
  \caption[]{\it The derivative function $N_r^D$ as  function of the rapidity $Y$ 
at $x_0=10^{-2}$. }
\label{scal_r}
\end{figure}

If the scaling behavior (\ref{SCALING}) takes place indeed, then both the ratios
$N_y^D/N_r^D$ and $\Re/N_r^D$ are $r_\perp$ independent functions. 
Let us first consider scaling with respect to $x$. Fig. \ref{scal_x} presents the derivatives 
 $N_y^D$ and $N_r^D$  as functions of the distance $r_\perp$ at fixed $x_0=10^{-2}$. 
Both functions  $N_y^D$ and $N_r^D$ have extrema placed at the same distance
depending on $x$. This is a consequence of the scaling behavior (\ref{SCALING}) and
equations (\ref{DX}) and (\ref{DR}). The extrema occur at certain $\tau_{max}$, such
that $\tilde N^{D\,\prime} (\tau_{max})=-\tau_{max} \tilde 
N^{D\,{\prime\prime}}(\tau_{max})$. In Fig.
\ref{scal_x},  $\tau_{max}$ is approached by varying $r_\perp$ at fixed $x$. Alternatively
it can be reached by varying $x$  at fixed  $r_\perp$ (Fig. \ref{scal_r}).

\begin{figure}[htbp]
\begin{tabular}{c c c c}
 \epsfig{file=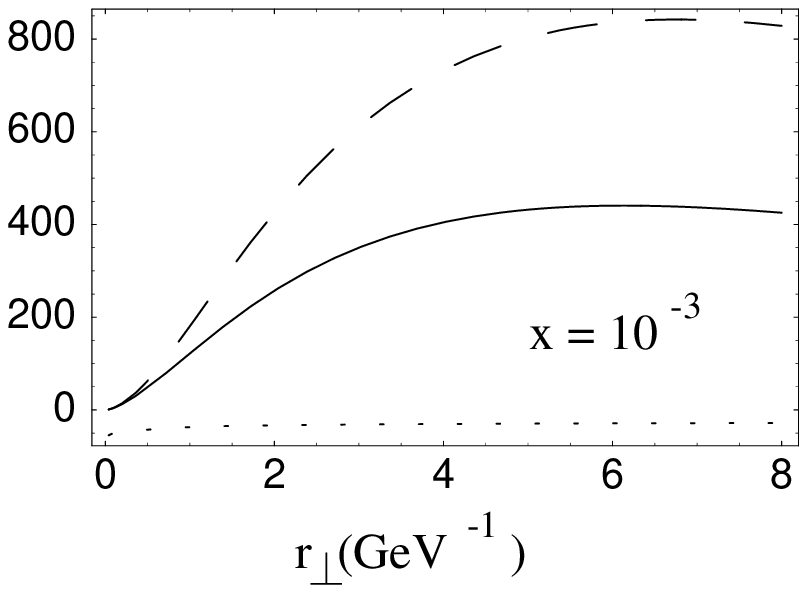,width=38mm, height=40mm}&
\epsfig{file=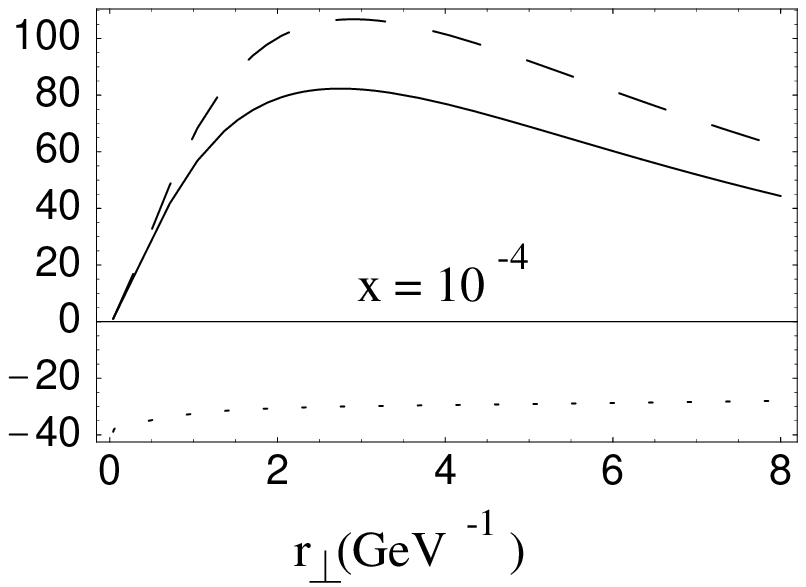,width=38mm, height=40mm}&
 \epsfig{file=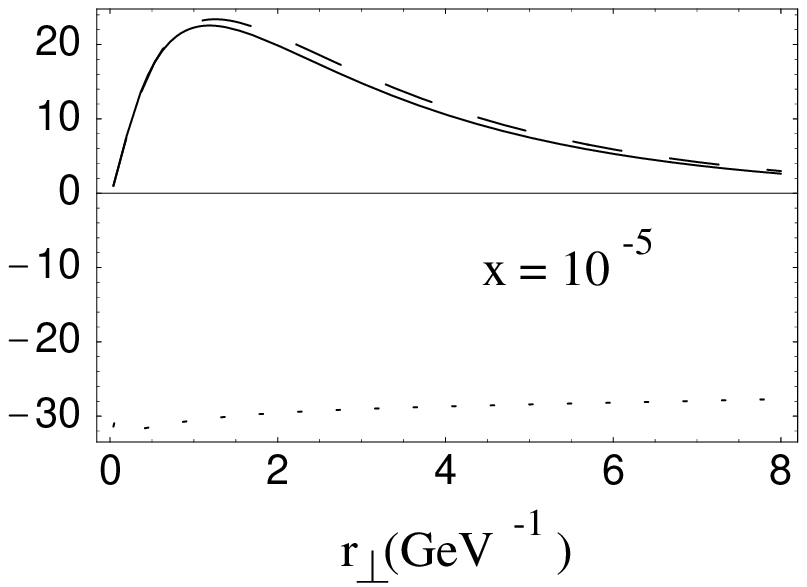,width=38mm, height=40mm}&
\epsfig{file=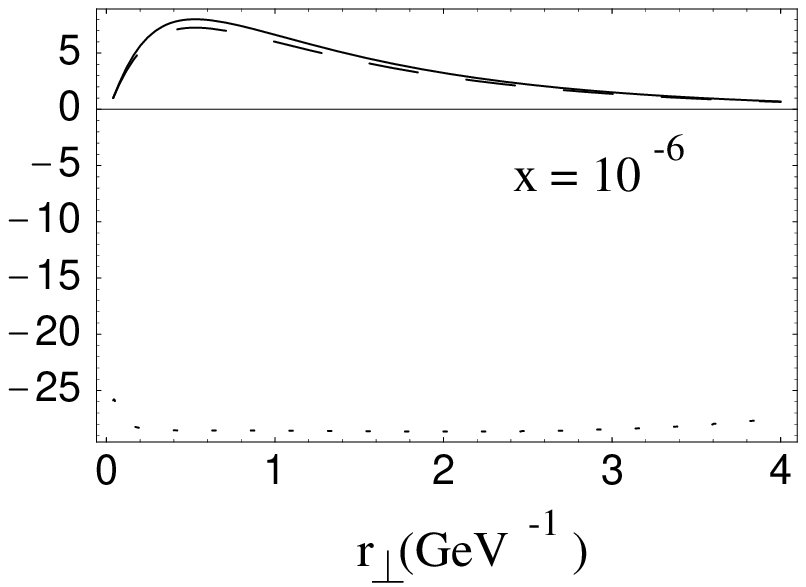,width=38mm, height=40mm}\\ 
\end{tabular}
  \caption[]{\it The  scaling   as a function of the distance  at fixed $x_0=10^{-2}$. 
The positive curves
  are $N_r^D/N_{r\,min}^D$ (dashed line) and   $N_y^D/N_{y\,min}^D$ (solid line). 
The dotted line is  $40\times R_a^D$. }
\label{ratio}
\end{figure}

Consider now the ratio function $R_a^D$:
\beq\label{Ra}
 R_a^D(r_\perp,x,x_0)\,\equiv\, \frac{ N_y^D}{ N_r^D}\,=\,\frac{\partial \ln (Q_s^D)^2}
{\partial \ln x}\,.
\eeq
If the scaling phenomenon takes place the function $R_a^D$ is expected to be $r_\perp$
independent. We study the scaling within the distance interval 
$0.04\, {\rm GeV^{-1}}\le r_\perp \le 10 \, {\rm GeV^{-1}}$ that corresponds to  
$0.25\, {\rm GeV^{2}}\le Q^2 \le 2.5\times 10^3\,  {\rm GeV^{2}}$.
Fig. \ref{ratio} presents the  results on the scaling. 
The three lines correspond to 
functions $N_r^D$ and $N_y^D$ 
divided by their minimal values within the interval, and the function
$R_a^D$ multiplied by the factor 40 to be seen on the scale.

The function $R_a^D$  is clearly observed to be a very slowly varying function of $r_\perp$ 
for all values of $x$ and $r_\perp$. Though at fixed $x$ 
the function $R_a^D$  cannot be  claimed to be exact constant, its variations with $r_\perp$ 
are very much suppressed comparing to the variations of the functions $N_r^D$ and $N_y^D$. 
For example, at $x=10^{-5}$ within the given interval the function $R_a^D$ changes by maximum
20\%, while  
within the very same interval both functions  $N_r^D$ and $N_y^D$ change in several
times. Then the relative fluctuation  is much less than 10\%, which  confirms the scaling. 
The phenomenon   holds with a few percent   accuracy  and it
 improves at smaller $x\simeq10^{-7}$ and in the  deep saturation
region. However to observe  this scaling behavior 
in these regions  is numerically more problematic since 
both derivatives $N_r^D$ and $N_y^D$  tend to zero. 

The above analysis was performed for the fixed value $x_0=10^{-2}$.
Within the errors the function $R_a^D\simeq - 0.75\pm 0.08$, 
 constant independent on both $r_\perp$ and $x$. Moreover, if we repeat the same program
but for different values of $x_0$ we discover quite similar scaling phenomena with
$R_a^D$ being numerically independent on $x_0$ as well. This observation implies  
\beq\label{lam}
Q_s^D(x,x_0)\,=\,Q_{s\,0}^D(x_0)\, x^{-\lambda}\,;\,\,\,\,\,\,\,\,\,\,\,\,\,\, \lambda\,=\,0.37\,\pm \,0.04\,.
\eeq
Note that the value obtained for $\lambda$ is in agreement with the one determined in the 
previous section.

\begin{figure}[htbp]
\begin{tabular}{c c }
 \epsfig{file=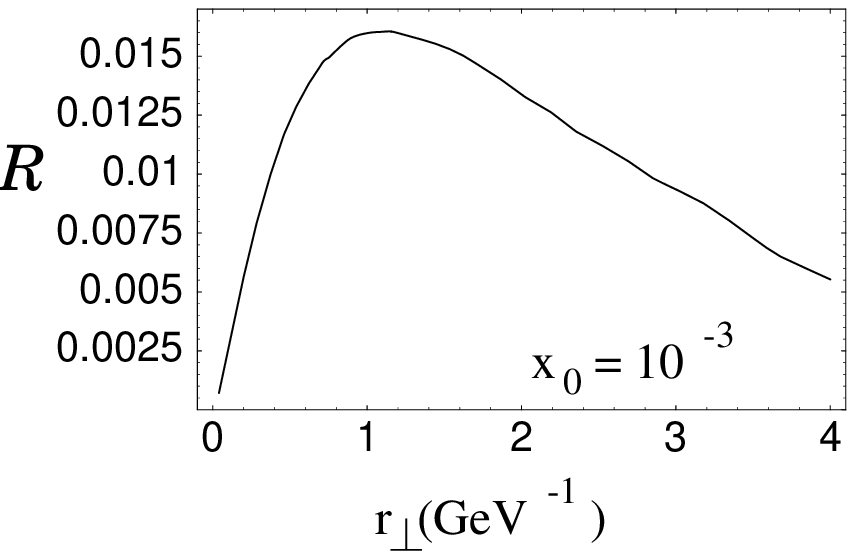,width=60mm, height=45mm}&
\epsfig{file=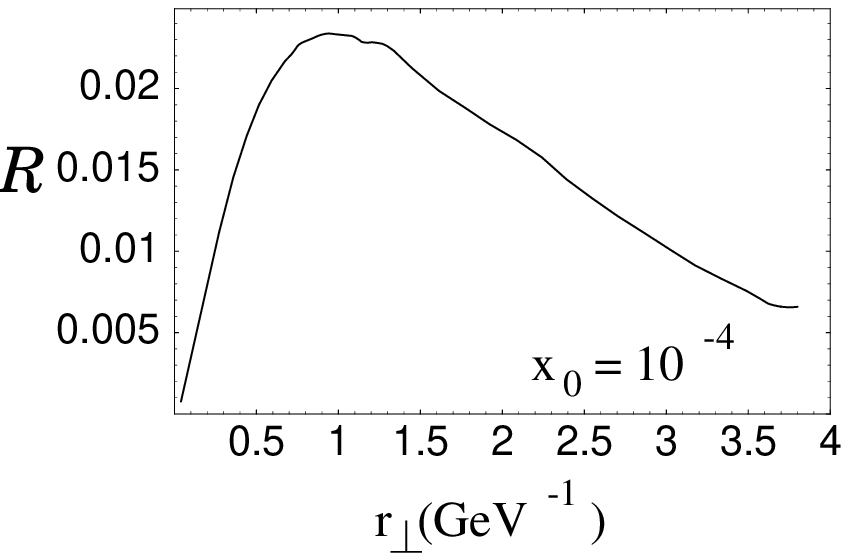,width=60mm, height=45mm}\\ 
\end{tabular}
  \caption[]{\it The  function $\Re$  versus distance at fixed  $Y=10$.  }
\label{Re}
\end{figure}

Let us now study the scaling behavior with respect to the variable $x_0$. To this goal
we investigate the function $\Re$ which is related to the $x_0$-dependence of the saturation
scale $Q_{s\,0}^D$. Assuming $Q_{s\,0}^D\sim x_0^\beta$ we predict  $\Re$ to have a maximum
at $\tau=\tau_{min}$. Fig. \ref{Re} displays the function $\Re$ as a function of the distance
at fixed $x=4.54\cdot 10^{-5}$ ($Y=10$).
In complete agreement with the scaling assumption (\ref{DX0}) the
 function $\Re$ possesses maximum with respect to $r_\perp$ variations. The
heights of the maxima are proportional to $\beta$.  Since $\tau\,\tilde N^{D\,\prime}(\tau)
|_{\tau=\tau_{max}}\simeq0.2$, $\beta$ can be estimated to be approximately $0.05\pm 0.02$
which agrees with the value deduced earlier. 

\begin{figure}[htbp]
\begin{tabular}{c c c c}
 \epsfig{file=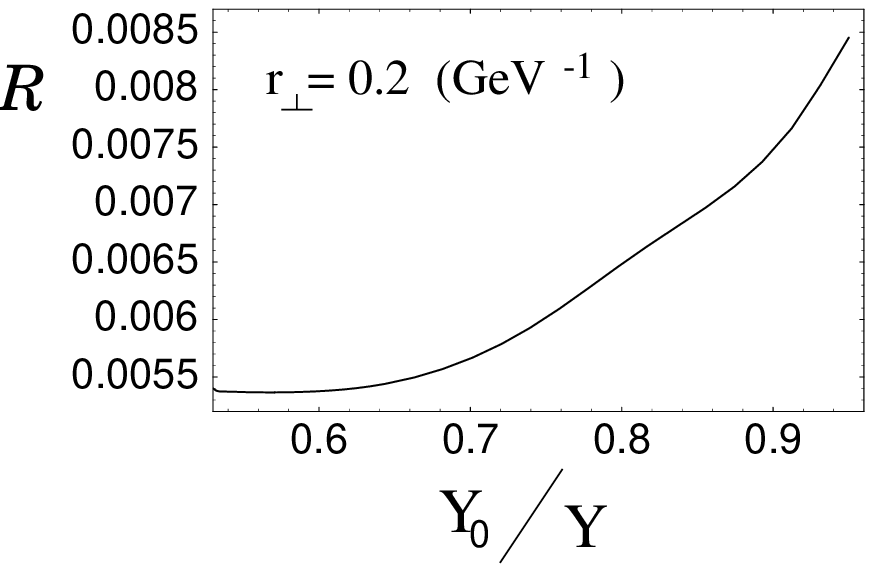,width=42mm, height=40mm}&
\epsfig{file=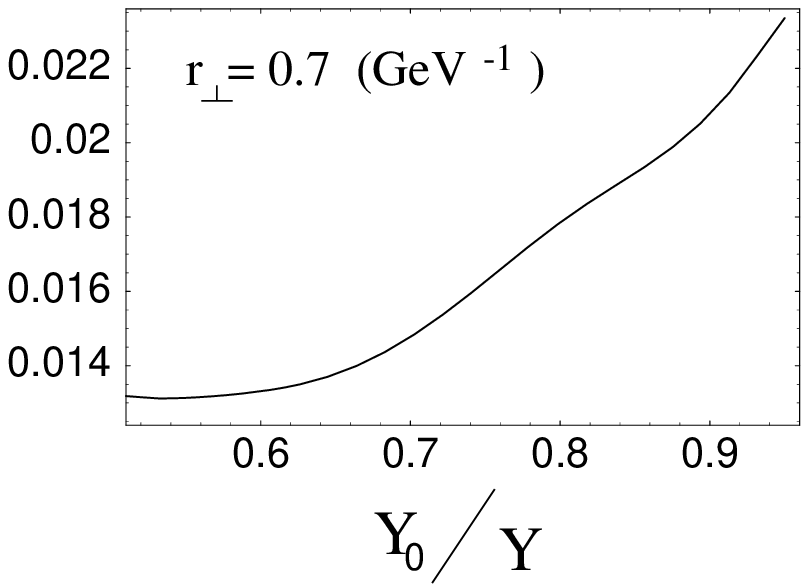,width=38mm, height=40mm}&
 \epsfig{file=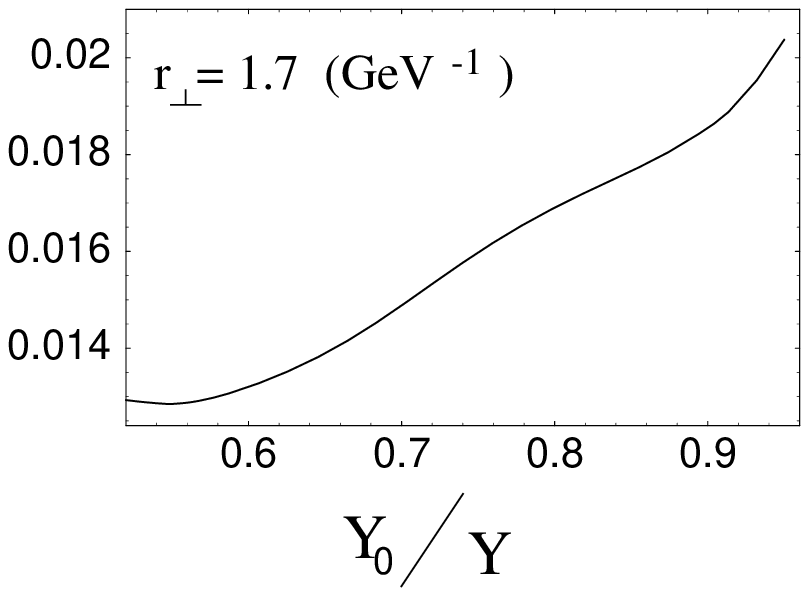,width=38mm, height=40mm}&
\epsfig{file=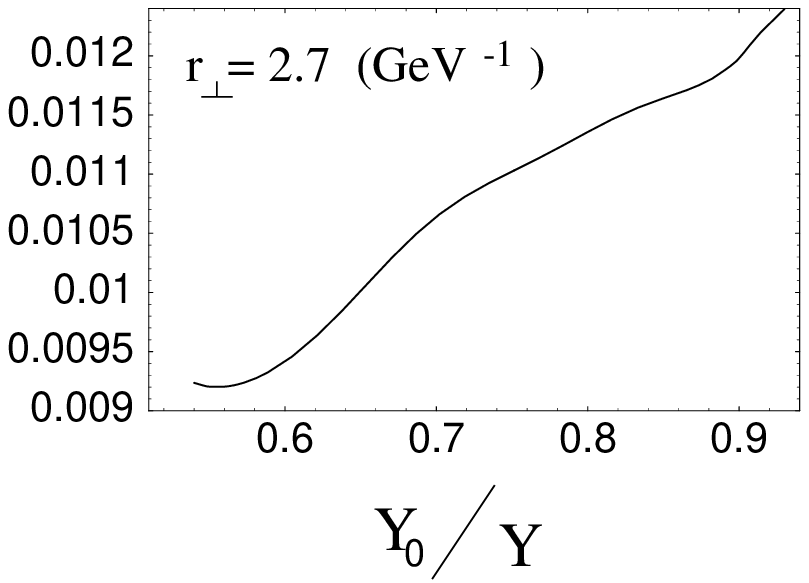,width=38mm, height=40mm}\\ 
\end{tabular}
  \caption[]{\it The function $\Re$ as  a function of the ratio $Y_0/Y$ 
at fixed $Y=10$. }
\label{Y0}
\end{figure}

We can learn more about the scaling if we consider the function 
$\Re$ as a function of $x_0$ or the energy gap $Y_0$. In Ref. \cite{LK} a model was built 
in which the function $\Re$ had a maximum with respect to $Y_0$ variation at fixed $Y$. 
We know now that this maximum is a consequence of the scaling phenomena. 
The dependence of $\Re$ on $Y_0$ at $Y=10$ is plotted in Fig. \ref{Y0}. 

No maxima is observed on the plots of  Fig. \ref{Y0}. In fact this is a sign of the scaling
violation so far avoided by the discussion. The scaling with respect to $x_0$ is not exact 
at $Y_0\simeq Y$.
Due to its smallness ($\Re\propto \beta$) the
function $\Re$ is most sensitive to small deviations from the scaling behavior:
\beq\label{dev}
\tilde N^D(r_\perp,x,x_0)\,=\,\tilde N^D_{scaling}(\tau)\,+\,\delta \tilde N^D(r_\perp,x,x_0)
\eeq
In the kinematic region of the investigation  variations of the function $\delta \tilde N^D$
with respect to $r_\perp$ and $x$ are small compared to variations of $\tilde N^D_{scaling}$. 
In contrary, the derivative of  $\delta N^D$ with respect to $Y_0$ is of the same order
as derivative of  $\tilde N^D_{scaling}$. This is the origin of the large errors of $\beta$ and 
the $x_0$ scaling violation at $x_0\simeq x$.

\begin{figure}[htbp]
 \epsfig{file=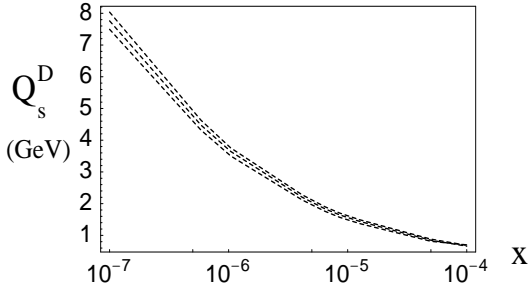,width=70mm, height=40mm}
\begin{minipage}{9.5 cm}
\vspace{-4.5cm}
 \caption[]{\it The saturation scale deduced from (\ref{defd}).
 The different curves correspond to  $x_0=10^{-2}$ 
(the upper curve), $10^{-3}$ (middle curve) and
 $x_0=10^{-4}$ (the lowest curve)}
\label{Qs_scal}
\end{minipage}
\end{figure}

In order to complete the analysis we propose yet another definition of the saturation 
scale based on the above presented scaling analysis. It is natural to define the saturation
radius at the position where  $\tau\,\tilde N^\prime(\tau)$ has maximum, namely at $\tau_{max}$:
\begin{itemize}
\item {\bf Definition (d):}
\beq \label{defd}
\left (\frac{\partial \,(\tau\,\tilde N^\prime(\tau))}
{\partial r_\perp^2 }\right )_{r_\perp^2=4/(Q_s^{D})^2}\,
=\,0\,.
\eeq
\end{itemize}
The saturation scale obtained from (\ref{defd}) is depicted in Fig. \ref{Qs_scal}. Note again
the weak dependence on the value of $x_0$.

\section{Conclusions}

The non-linear evolution equation (\ref{DDEQ}) is solved numerically by the method of iterations.
The solutions obtained are in agreement with the unitarity constraints: the diffraction dissociation
is larger than just the elastic scattering but smaller or equal than half of the total.

The diffractive saturation scale $Q_s^D$ is estimated form the solutions of  (\ref{DDEQ}) basing
on four different definitions of the saturation scale. Though  there exists a significant uncertainty in
the absolute values of the scale its $x$-dependence is found to be  the same as of  
$Q_s$ - saturation scale deduced from the non-linear equation for $\tilde{N}$ \cite{BA,KO}. In fact 
this result is quite natural.
The dependence of the saturation scale on $x$ is entire property of the evolution equation and it
should not depend on both initial conditions and  saturation scale definition. 
The  
saturation scale $Q_s^D$ is discovered to be almost independent on the minimal gap $x_0$.

The scaling phenomena with respect to all variables were studied in details. The scaling 
with respect to  $x$ is well established. It holds with a few percent accuracy in the whole 
kinematic region  investigated. The discovered scaling should manifest itself in the experiments
 on diffraction, and hence it would be interesting to search for it in the 
$F^D_2(x,Q^2)/(Q^2\,S)$ experimental data ($S$ stands for the target transverse area).

The numerically observed small scaling violation shows up when we consider  
the scaling  with respect to $x_0$. This happens due to the weak sensitivity of the solutions to the variation
of $x_0$. As a result, the variations of the solutions with respect to $x_0$ are of the same order as the 
scaling violation. The scaling  sets in at $x\ll x_0$ but is violated at $x\sim x_0$.

The detailed analysis of the ratio between the total diffractive 
dissociation and the total DIS cross
section will be presented in a separate publication \cite{LL1}. 
Our preliminary computations show that this ratio
happens to be independent on the central mass energy in agreement with the
experimental data \cite{ZEUSDATA}. 
This independence can be traced back
to the scaling property displayed by the amplitudes $N$ and $N^D$  
and to the fact that both saturation
scales depend on $x$ with the very same power $\lambda$.

\section*{Acknowledgments}

The authors are very much indebted to Jochen Bartels,
Krystoff Golec-Biernat and Yuri Kovchergov
 for numerous helpful discussions about   
diffraction production in DIS. We would like to thank 
 Asher Gotsman,  Uri Maor, Eran Naftali and Kirill Tuchin
 for many
informative and encouraging discussions. We thank DESY theory group 
and Hamburg University Institute of theoretical physics  for their 
hospitality and
creative atmosphere during several stages of this work.

 The research of 
E. L. was supported in part by the BSF grant $\#$ 9800276, by GIF grant $\#$ I-620.-22.1411444  and by
Israeli Science Foundation, founded by the Israeli Academy of Science
and Humanities.
The work of M.L. was partially supported by the Minerva Foundation and its
financial  help is gratefully acknowledged.

\end{document}